\documentclass{jps-cp}
\usepackage{txfonts} 

\usepackage{graphicx,color}
\usepackage{amsmath,amssymb,bm,ascmac,calc}
\usepackage{wrapfig}

\title{Deformability of neutron-deficient Sn nuclei
  studied with semi-realistic interaction}

\author{H. \textsc{Nakada}$^{1,2}$, Y. \textsc{Omura}$^{3}$,
  K. \textsc{Abe}$^{3}$, M. \textsc{Takahashi}$^{1}$
  and T. \textsc{Inakura}$^{4}$}

\inst{
  $^{1}$Department of Physics, Graduate School of Science, Chiba University,
  Inage, Chiba 263-8522, Japan\\
  $^{2}$Research Center for Nuclear Physics, Osaka University,
  Ibaraki, Osaka 567-0047, Japan\\
  $^{3}$Department of Physics, Graduate School of Science and Engineering, Chiba University,
  Inage, Chiba 263-8522, Japan\\
  $^{4)}$\,Office of Institute Strategy, Institute of Science Tokyo,
  Meguro, Tokyo 152-8550, Japan
}

\email{nakada@faculty.chiba-u.jp}

\recdate{September 8, 2026}

\abst{
  Although Sn nuclei have long been considered typical spherical nuclei,
  the $E2$ transitions in neutron-deficient Sn nuclei have posed a puzzle.
  We argue that their deformability may resolve this issue,
  based on self-consistent calculations with a semi-realistic interaction.
  The significant role of the pseudo-spin symmetry,
  which is attributed to the effects of the tensor force, is also discussed.
}

\kword{HFB\,+\,QRPA, CHFB\,+\,AMP, Sn nuclei, $E2$ transition,
  pseudo-spin symmetry, tensor force}

\begin{document}
\maketitle


\section{Introduction}

Sn nuclei have long been considered typical spherical nuclei.
They exhibit an almost constant $E_x(2_1^+)$ in $52\leq N\leq 80$
and parabolic behavior of $B(E2;0^+_1\to 2^+_1)$ in $N\gtrsim 66$,
suggesting that they are excellent examples
of the seniority scheme~\cite{Talmi}.
However, recent measurements of $B(E2;0^+_1\to 2^+_1)$ in $N\lesssim 60$
have posed a puzzle.
Significantly deviating from the parabolic behavior
predicted by the seniority picture,
these data cannot be reproduced by the conventional shell model
within a single major shell~\cite{Doo14};
the measured $B(E2)$ values are stronger
than predicted by the conventional shell model
with the neutron effective charge adjusted to the $N\gtrsim 66$ data.
Furthermore, calculations based on
self-consistent mean-field (SCMF) theories,
primarily the quasiparticle random-phase approximation (QRPA)
on top of Hartree-Fock-Bogoliubov (HFB) solution,
overestimate the experimental data~\cite{CTP12,PM14,Ans05}.
Thus, these SCMF results not only fail to reproduce the data,
but also yield discrepancies opposite to those of the shell model.

We have reinvestigated the $B(E2)$ values in Sn nuclei
within the SCMF framework,
and how the aforementioned puzzle can be resolved~\cite{ONAT23}.
In the SCMF calculations including HFB\,+\,QRPA,
the only non-trivial input is the effective nucleonic interaction,
for which we adopt the semi-realistic interaction M3Y-P6~\cite{Nak13,Nak20}.
No external parameters, such as effective charges, are necessary
for the $E2$ operator.
A shell-model calculation in an extended model space
has indicated that proton excitation across the $Z=50$ magic number
plays a key role in $B(E2)$ of neutron-deficient Sn nuclei~\cite{Tog18}.
It is of interest to elucidate whether and how
this indication can be compatible with the SCMF results.

\section{Numerical setups}

We implement self-consistent calculations
with a semi-realistic nucleonic interaction;
specifically, spherical HFB\,+\,QRPA
and axial constrained-HFB (CHFB) calculations
on top of which angular-momentum-projection (AMP) is carried out.

In SCMF calculations, including QRPA and AMP,
the effective Hamiltonian is given
by $\hat{H} = \hat{K} + \hat{V}_N + \hat{V}_{\mathrm{Coul}}
- \hat{H}_{\mathrm{c.m.}}$,
where $\hat{K}$, $\hat{V}_{\mathrm{Coul}}$, and $\hat{H}_{\mathrm{c.m.}}$
are the non-relativistic kinetic energy, the Coulomb interaction,
and the center-of-mass term, respectively.
The effective nucleonic interaction $\hat{V}_{N}$ consists of
central (density-independent and dependent terms), LS, and tensor channels,
to which the semi-realistic nucleonic interaction M3Y-P6~\cite{Nak13,Nak20}
is applied throughout this paper.
Based on the M3Y-Paris interaction derived from the $G$-matrix~\cite{M3Y},
this interaction is connected to bare nucleonic interaction,
with minimal modification so as to reproduce nuclear saturation
and $\ell s$ splitting~\cite{Nak03}.
It is free from most unphysical instabilities~\cite{DPJ21}
and has been confirmed to have significant advantages
in describing shell structures~\cite{NS14}.
Note that the tensor force is untouched from the original M3Y-Paris interaction,
and the longest-range part of the central channel
is the one-pion exchange potential itself.
Numerical calculations have been implemented
using the code \texttt{MFGSB}~\cite{mfgsb}
and its extensions~\cite{NMYM09,AN22}.

\section{Results}

\begin{figure}[tbh]
\includegraphics[scale=0.6]{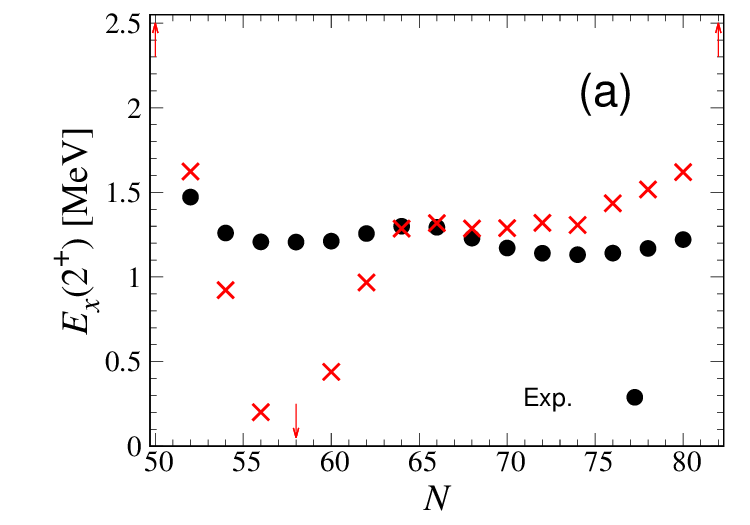}
\includegraphics[scale=0.6]{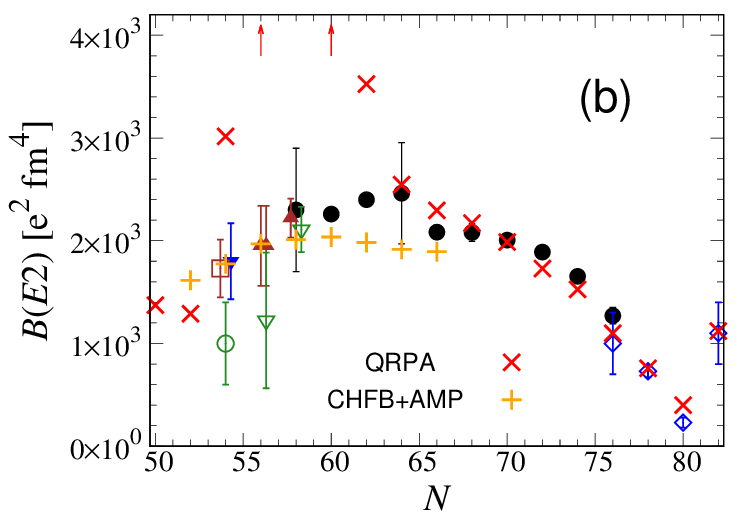}
\caption{
(a) $E_x(2_1^+)$ and (b) $B(E2;0^+_1\to 2^+_1)$ in $^{100-132}$Sn.
The spherical HFB\,+\,QRPA results (red crosses)
and the reference $B(E2)$ values obtained by the CHFB\,+\,AMP calculations
(orange pluses) for $^{102-116}$Sn
are compared with the experimental data (black and colored circles,
diamonds, triangles, and a square)~\cite{Doo14,NuDat,
  Rad05,Eks08,Gua13,Bad13,Sic20},
Quoted from Ref.~\cite{ONAT23}.
}
\label{HFB+QRPA}
\end{figure}

The results of the spherical HFB\,+\,QRPA calculations
are presented in Fig.~\ref{HFB+QRPA},
together with reference values from CHFB\,+\,AMP
that will be explained later.
We find that $E_x(2^+_1)$ and $B(E2;0^+_1\to 2^+_1)$
are excellently reproduced in $N\geq 64$,
with no adjustable parameters.
However, underestimations of $E_x(2^+)$ and overestimations of $B(E2)$
are obtained in $54\leq N\leq 62$,
indicating excessively strong quadrupole collectivity.
~\vspace*{-3mm}

\vspace*{-3mm}

\begin{wrapfigure}{r}{6.7cm}
\includegraphics[scale=0.55]{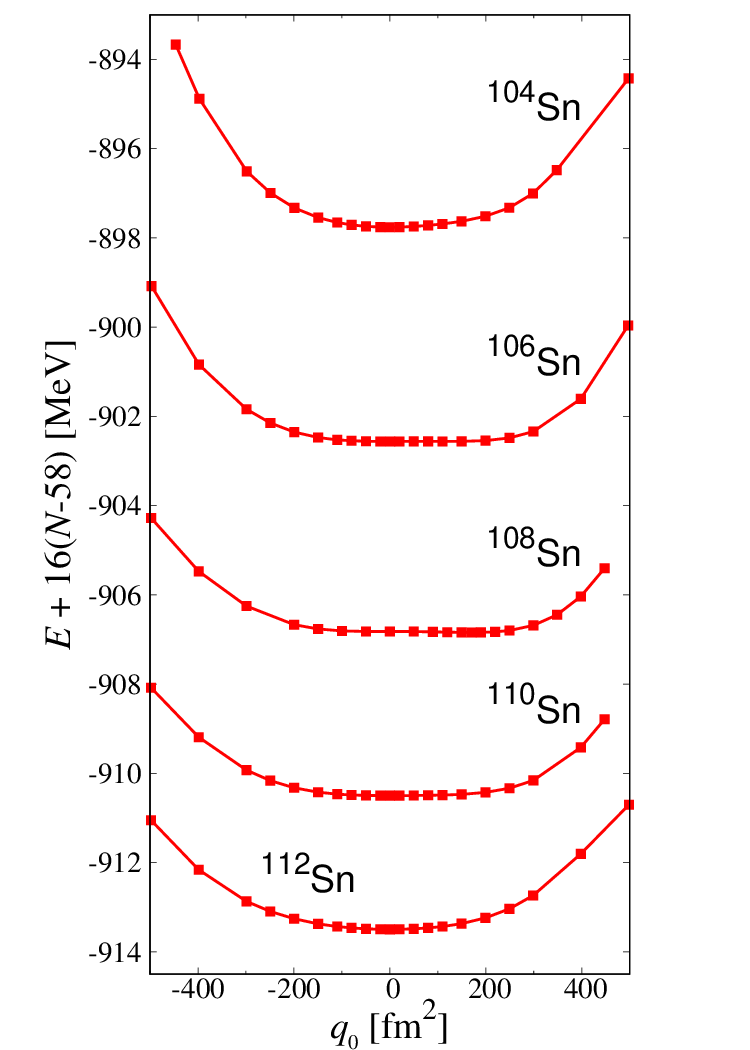}
\caption{
Potential energy curves for $^{104-112}$Sn.
}
\label{CHFB}
\end{wrapfigure}

To investigate the mechanisms in $54\leq N\leq 62$,
we have carried out constrained HFB (CHFB) calculations,
by constraining the mass quadrupole moment $q_0$~\cite{SNM16}.
The potential energy curves (PECs) are depicted in Fig.~\ref{CHFB}.
The PECs are surprisingly flat around $N=58$,
demonstrating a deformability and accounting for the failure of the QRPA.
It is noted that this flatness remains
within a narrow range of the deformation parameter, $|\beta_2|<0.08$.

To examine whether this deformability in the small $\beta_2$ range
can account for the measured $E2$ strengths,
we have further implemented the angular-momentum projection (AMP)~\cite{AN22}
on top of the CHFB wave functions.
The $B(E2)$ values obtained at the CHFB state with the lowest energy after AMP
are also plotted in Fig.~\ref{HFB+QRPA}.
Since the flat PECs suggest that a superposition of the CHFB states
is desirable,
which is left for a future work,
these values serve merely as a reference.
Nevertheless, we observe that the measured $B(E2)$ values can be
described well by this weak deformation.
This consequence appears consistent with that of the shell model
in an extended model space~\cite{Tog18}.

\section{Roles of pseudo-spin symmetry and tensor force}

The flat PEC arises from a delicate balance between the shell structure
and pairing correlations.
As discussed in Ref.~\cite{ONAT23},
the $B(E2)$ values and relevant quantities can hardly be reproduced
simultaneously,
as long as the neutron-deficient Sn nuclei stay spherical.
The deformability is a consequence
of the pseudo-spin symmetry (PSS)~\cite{AHS69,HA69}
between the $n0g_{7/2}$ and $n1d_{5/2}$ orbitals,
which allows neutrons to occupy these two orbits
with almost equal probabilities.

\begin{figure}[tbh]
\begin{center}
\includegraphics[scale=0.8]{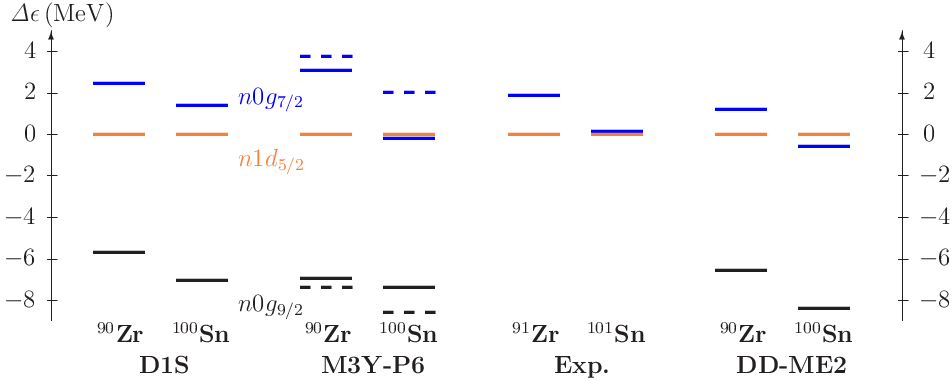}
\caption{
S.p. energies relative to $n1d_{5/2}$ at Zr and Sn,
for several SCMF results~\cite{D1S,DD-ME2} and those extracted
from experimental data~\cite{NuDat,AME2020}.
M3Y-P6 results obtained after subtracting the contribution of the tensor force
are also displayed (dashed lines).
Quoted from Ref.~\cite{NI24}.
}
\label{PSS}
\end{center}
\end{figure}

As clarified in Ref.~\cite{NI24}, 
a notable feature of the PSS is its $Z,N$-dependence.
Concerning the $50<N<64$ region,
$N=56$ behaves as a magic number in Zr isotopes~\cite{NS14,MN18},
implying a significant breaking of the PSS,
in contrast to the presence of PSS in Sn that gives rise to the deformability.
This $N$-dependence is attributed to the effects
of the tensor force~\cite{NS14,Ots05}.
Because the M3Y-P6 interaction contains a realistic tensor force,
it properly describes both the PSS in Sn and its breaking in Zr,
as shown in Fig.~\ref{PSS}.
It is mentioned that flat PECs were predicted in Sn
by some interactions without tensor force (\textit{e.g.}, D1S),
but in a shifted region of $N$~\cite{PM14},
owing to the shell structure
that is not compatible with the observed PSS.

The PSS has often been discussed in the context
of relativistic symmetry~\cite{Gin97,LMZ15}.
Interestingly, similar $Z,N$-dependence of the PSS is predicted
by the relativistic framework without tensor-force effects
and the present non-relativistic calculation with the realistic tensor force
in many cases.
However, an exception has been found
in the variation of the $p0d_{3/2}$-$p1s_{1/2}$ levels
from $^{40}$Ca to $^{34}$Si,
where available experimental data seem to favor the tensor-force mechanism
for the PSS~\cite{NI24}.

\section{Summary}

The low-energy quadrupole collectivity in Sn nuclei,
$E_x(2^+_1)$ and $B(E2;0^+_1\to 2^+_1)$,
has been studied through self-consistent calculations
with the semi-realistic interaction M3Y-P6.
Without artificial parameters,
the spherical HFB\,+\,QRPA results are in good agreement
with the experimental data for $N\gtrsim 64$.
Implementing the axial CHFB calculations,
we find remarkably flat PECs within the range $|\beta_2|<0.08$
for neutron-deficient Sn nuclei,
demonstrating their deformability and accounting for the failure of the QRPA.
This flat PEC takes place due to the balance
between the PSS with respect to the $n0g_{7/2}$ and $n1d_{5/2}$ orbitals
and the pairing correlations.
The AMP on top of the CHFB wave functions indicates
that this deformability may resolve the puzzle,
although a superposition of CHFB wave functions is desired
for a complete understanding.

\section*{Acknowledgment}

This work is supported in part by JSPS KAKENHI, Grant No. JP24K07012.
The numerical calculations were performed
using Oakforest PACS and Wisteria/BDEC,
under the MCRP Program at CCS, University of Tsukuba,
Yukawa-21 at YITP, Kyoto University,
and HITAC SR24000 at CUDTEC, Chiba University.


\begin{thebibliography}{9}
\bibitem{Talmi} I. Talmi,
  \textit{Simple Models of Complex Nuclei}, (Harwood, Chur, 1993).
\bibitem{Doo14} P. Doornenbal, \textit{et al.},
  Phys. Rev. C \textbf{90}, 061302(R) (2014).
\bibitem{CTP12} B.G.~Carlsson, J.~Toivanen, A.~Pastore,
  Phys. Rev. C \textbf{86}, 014307 (2012).
\bibitem{PM14} S. P\'{e}ru, M. Martini,
  Eur. Phys. J. \textbf{A} 50, 88 (2014).
\bibitem{Ans05} A. Ansari, Phys. Lett. \textbf{B} 623, 37 (2005).
\bibitem{ONAT23} Y. Omura, \textit{et al.},
  Phys. Rev. C \textbf{108}, 054308 (2023).
\bibitem{Nak13} H. Nakada, Phys. Rev. C \textbf{87}, 014336 (2013).
\bibitem{Nak20} H. Nakada, Int. J. Mod. Phys. E \textbf{29}, 1930008 (2020).
\bibitem{Tog18} T.~Togashi, \textit{et al.},
  Phys. Rev. Lett. \textbf{121}, 062501 (2018).
\bibitem{M3Y} N. Anantaraman, H. Toki, G.F. Bertsch,
  Nucl. Phys. \textbf{A} 398, 269 (1983).
\bibitem{Nak03} H. Nakada, Phys. Rev. C \textbf{68}, 014316 (2003).
\bibitem{DPJ21} D.~Davesne, A.~Patore, J.~ Navarro,
  Prog. Part. Nucl. Phys. \textbf{120}, 103870 (2021).
\bibitem{NS14} H. Nakada, K. Sugiura,
  Prog. Theor. Exp. Phys. \textbf{2014}, 033D02.
\bibitem{mfgsb} H. Nakada, Chiba University Repository, DOI: 10.20776/900123722\\
  (https://opac.ll.chiba-u.jp/da/curator/900123722/?lang=1).
\bibitem{NMYM09} H.~Nakada, \textit{et al.},
  Nucl. Phys. A \textbf{828}, 283 (2009).
\bibitem{AN22} K. Abe, H. Nakada, Phys. Rev. C \textbf{106}, 054317 (2022);
  J. Phys. G \textbf{51}, 035101 (2024).
\bibitem{NuDat} http://www.nndc.bnl.gov/nudat3/.
\bibitem{Rad05} D.C.~Radford, \textit{et al.}
  Nucl. Phys. A \textbf{752}, 264c (2005).
\bibitem{Eks08} A.~Ekstr\"{o}m, \textit{et al.},
  Phys. Rev. Lett. \textbf{101}, 012502 (2008).
\bibitem{Gua13} G.~Guastalla, \textit{et al.},
  Phys. Rev. Lett. \textbf{110}, 172501 (2013).
\bibitem{Bad13} V.M.~Bader, \textit{et al.},
  Phys. Rev. C \textbf{88}, 051301(R) (2013).
\bibitem{Sic20} M.~Siciliano, \textit{et al.},
  Phys. Lett. B \textbf{806}, 135474 (2020).
\bibitem{SNM16} Y.~Suzuki, H.~Nakada, S.~Miyahara,
  Phys. Rev. C \textbf{94}, 024343 (2016).
\bibitem{AHS69} A. Arima, M. Harvey, K. Shimizu,
  Phys. Lett. \textbf{B} 30, 517 (1969).
\bibitem{HA69} K.T. Hecht, A. Adler,
  Nucl. Phys. \textbf{A} 137, 129 (1969).
\bibitem{NI24} H. Nakada, T. Inakura, Phys. Rev. C \textbf{110}, L051301 (2024).
\bibitem{MN18} S. Miyahara, H. Nakada, Phys. Rev. C \textbf{98}, 064318 (2018).
\bibitem{Ots05} T. Otsuka, \textit{et al.},
  Phys. Rev. Lett. \textbf{95}, 232502 (2005).  
\bibitem{Gin97} J.N. Ginocchio, Phys. Rev. Lett. \textbf{78}, 436 (1997).
\bibitem{LMZ15} H. Liang, J. Meng, S.-G. Zhou,
  Phys. Rep. \textbf{570}, 1 (2015).
\bibitem{D1S} J.F. Berger, M. Girod, and D. Gogny,
  Comp. Phys. Commun. \textbf{63}, 365 (1991).
\bibitem{DD-ME2} G.A. Lalazissis, \textit{et al.},
  Phys. Rev. C \textbf{71}, 024312 (2005).
\bibitem{AME2020} M. Wang, \textit{et al.},
  Chin. Phys. C \textbf{45}, 030003 (2021).

\end{thebibliography}
\end{document}